\documentclass[aps,floats,showpacs,nofootinbib,floatfix]{revtex4}
\usepackage{epsfig}
\usepackage{amsfonts}
\usepackage{amsmath}
\usepackage{graphicx}
\usepackage{amssymb}
\begin{document}
\title{Model analysis of the world data on the pion transition form factor}
\author{S. Noguera}
\email{Santiago.Noguera@uv.es}
\author{V. Vento}
\email{Vicente.Vento@uv.es}
\affiliation{Departamento de Fisica Teorica and Instituto de F\'{\i}sica Corpuscular,
Universidad de Valencia-CSIC, E-46100 Burjassot (Valencia), Spain.}
\date{\today }

\begin{abstract}
We discuss the impact of recent Belle data on 
our description of  the pion transition form factor 
based on the assumption that a perturbative formalism
and a nonperturbative one can be matched in a physically acceptable manner at a
certain hadronic  scale $Q_{0}$. We discuss the implications of the different parameters of the model
in comparing with world data and conclude that 
within experimental errors our description remains valid.  Thus we can assert
that the low $Q^2$ nonperturbative description together with an additional $1/Q^2$ term at the matching scale have a strong influence on the $Q^2$
behavior up to very high  values of $Q^2$ .

\end{abstract}

\maketitle

New data of the pion transition form factor ($\pi TFF$) from the Belle collaboration have just appeared
 \cite{Uehara:2012ag}. These data, above 10 GeV$^2$, 
are smaller in magnitude than the previous BABAR data \cite{Aubert:2009mc}, which generated considerable excitement.
The question to unveil is the scale of  asymptotia. BABAR data, taken at face value, implied that asymptotic QCD behavior lies at much higher 
$Q^2$ than initially  expected \cite{Lepage:1980fj,Chernyak:1983ej}.  Belle data seem to lower that scale.
We show here that our scheme can accomodate easily all data without changing the physical input.
 
At the time of the BABAR data we  developed a formalism to calculate  the $\pi TFF$ \cite{Noguera:2010fe},
which consists of three ingredients: \textit{\ i}) a
low energy description of the $\pi TFF$; \textit{\ ii}) a high energy
description of the $\pi TFF$; \textit{\ iii}) a matching condition between the
two descriptions at a scale $Q_{0}$ characterizing the separation between the
two regimes. For the low energy description we took a parametrization
of the low energy data to avoid model dependence at $Q_{0}$. 
The high energy description of the $\pi TFF$, defined by the pion
Distribution Amplitude ($\pi DA$), contains Quantum Chromodynamic (QCD)
evolution from $Q_{0}$ to any higher $Q$, a  mass cut-off
to make the formalism finite, and  an additional $1/Q^2$  term which  leads to  modifications of the matching condition.

Let us recall some aspects of the formalism.
The high energy description, to lowest order in perturbative QCD, for the transition form factor in the process
$\pi^{0}\rightarrow\gamma\,\gamma^{\ast}$ in terms of the pion distribution amplitude ($\pi DA$), is given by 

\begin{equation}
Q^{2}F(Q^{2})=\frac{\sqrt{2}f_{\pi}}{3}\int_{0}^{1}\frac{dx}{x+\frac{M^{2}%
}{Q^{2}}}\phi_{\pi}(x,Q^{2}).
\label{tff_R}
\end{equation}
We follow the proposal of Polyakov \cite{Polyakov:2009je} and
Radyushkin 
\cite{Radyushkin:2009zg} and introduce a cutoff mass $M$ to make the expression finite.
$Q^{2}=-q^{2},$ $q_{\mu}$ is the momentum of the virtual photon,
$\phi_{\pi}\left(  x,Q^{2}\right)  $ is $\pi DA$ at the $Q^{2}$ scale and $f_\pi =0.131$ GeV. 
In this expression, the $Q^{2}$ dependence appears through the QCD evolution of the
$\pi DA$. 

Despite the fact that several models reproduce the low energy data, in order
to have a model independent expression for the form factor at low virtualities, we adopted a
monopole parametrization of the $\pi TFF$ in the low energy region as 
\begin{equation}
F^{LE}\left(  Q^{2}\right)  =\frac{F\left(  0\right)  }{1+a\frac{Q^{2}}%
{m_{\pi^{0}}^{2}}}\,. \label{tff_EXP}
\end{equation}
with $F\left(  0\right)  =0.273(10)$ GeV$^{-1}$ and $a=0.032\left(  4\right) $ \cite{Nakamura:2010zzi},  determined
from the experimental study of $\pi^{0}\rightarrow\gamma\,e^{+}\,e^{-}$ \cite{Amsler:2008zzb}. 

\begin{figure}[htb]
\begin{center}
\includegraphics[scale=0.5]{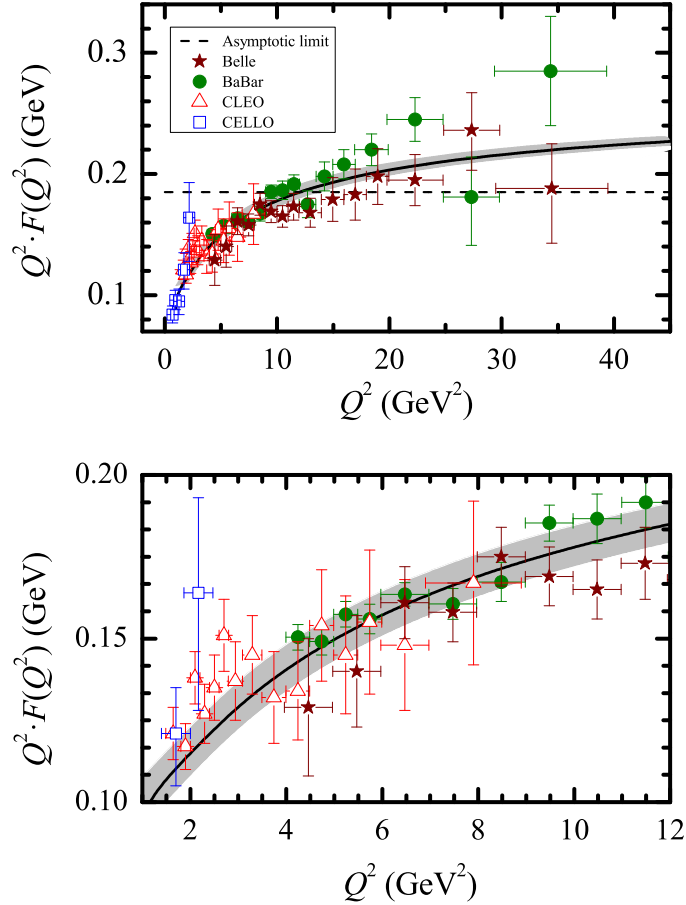}
\end{center}
\vskip -1.3cm
\caption{We show the result for the transition form factor  in our formalism for 
$M=0.690$ GeV, $a=0.032$ and $C_{3}=2.98\,10^{-2}$ GeV$^{3}$ and defining the matching point at $Q_{0}=1$ GeV (solid line).
The band region results from the indeterminacy in $\Delta a = \pm 0.004$. The lower plot shows the detailed behavior for low virtuality. Data are taken from CELLO \cite{Behrend:1990sr},  CLEO \cite{Gronberg:1997fj}, BABAR \cite{Aubert:2009mc} and Belle \cite{Uehara:2012ag}.}
\label{Fig1}

\end{figure}

\begin{figure}[htb]
\begin{center}
\includegraphics[scale=0.5]{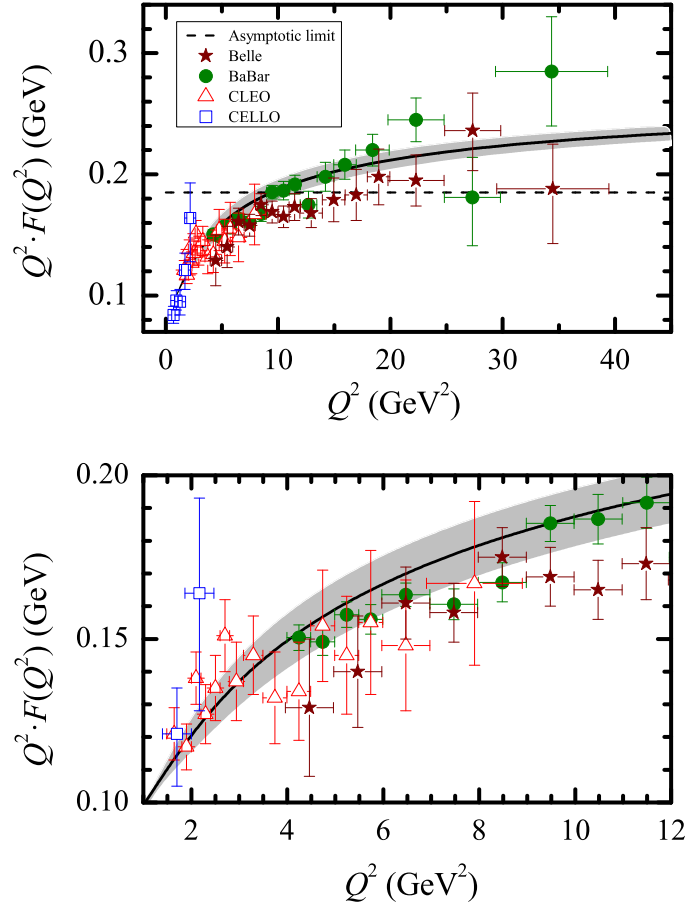}
\end{center}
\vskip -1.3cm
\caption{We show the result for the transition form factor  in our formalism for 
$M=0.620$ GeV, $a=0.032$ and the value of $C_{3}= 1.98\,10^{-2}$ GeV$^3$ corresponding to $20\%$ of the contribution at the matching point at $Q_{0}=1$ GeV (solid line).
The band  region gives the
variation of the results due in $ \pm 10 \%$ in the contribution of higher twist. The lower plot shows the detailed behavior for low virtuality. Data are taken from CELLO \cite{Behrend:1990sr},  CLEO \cite{Gronberg:1997fj}, BABAR \cite{Aubert:2009mc} and Belle \cite{Uehara:2012ag}.}
\label{Fig2}
\end{figure}

Additional power corrections can be introduced in Eq. \ref{tff_R} by adding to the
lowest order calculation a term proportional to $Q^{-2},$
\begin{equation}
Q^{2}F(Q^{2})=\frac{\sqrt{2}f_{\pi}}{3}\int_{0}^{1}\frac{dx}{x+\frac{M^{2}%
}{Q^{2}}}\phi_{\pi}(x,Q^{2})+\frac{C_{3}}{Q^{2}}. \label{tff_R_T3}%
\end{equation}

Using a constant $\pi$ DA  the matching condition becomes \cite{Noguera:2010fe},

\begin{equation}
\frac{\sqrt{2}f_{\pi}}{3}ln\frac{Q_{0}^{2}+M^{2}}{M^{2}}+\frac{C_{3}}%
{Q_{0}^{2}}=\frac{F\left(  0\right)  \,Q_{0}^{2}}{1+a\frac{Q_{0}^{2}}%
{m_{\pi^{0}}^{2}}}, \label{cont3}%
\end{equation}
with $Q_{0}=1$ GeV.  This equation allows to determine $M,$ once we have fixed the value of
$C_{3}.$

\begin{figure}[htb]
\begin{center}
\includegraphics[scale=0.5]{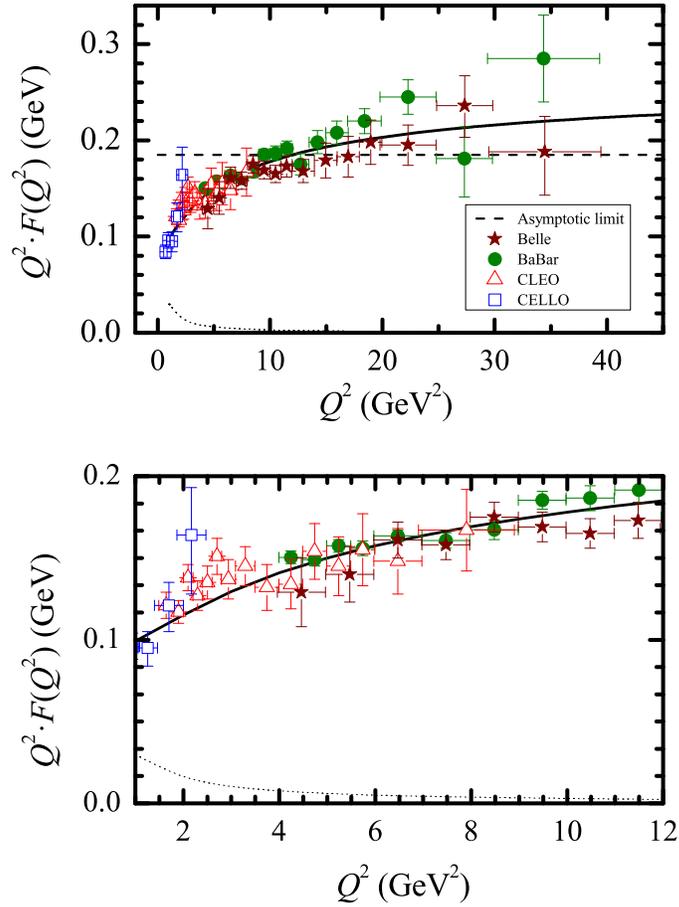}
\end{center}
\vskip -1.3cm
\caption{We show the result for the transition form factor  in our formalism for 
$M=0.690$ GeV, $a=0.032$ and the value of $C_{3}= 2.98\,10^{-2}$ GeV$^3$ corresponding to 
$30\%$ of the contribution at the matching point at $Q_{0}=1$ GeV (solid line).The lower plot shows the detailed behavior for low virtuality.The dotted curve represents
the higher twist contribution.Data are taken from CELLO \cite{Behrend:1990sr},  CLEO \cite{Gronberg:1997fj}, BABAR \cite{Aubert:2009mc} and Belle \cite{Uehara:2012ag}.}
\label{Fig3}
\end{figure}

We analyze here the sensitivity of the data to the various parameters involved.  We keep as close as possible to our previous fit analyzing the data with respect to small variations in the low virtuality parameter $a$ and in the higher twist parameter $C_3$. In Fig.\ref{Fig1} we show
the effect of the precision in the determination of the monopole parametrization. We see that as $a$ increases from $0.032$ to $0.036$, i.e. within the error bars, 
the $\pi TFF$ decreases. The sensitivity to $C_3$  is shown in Fig. \ref{Fig2} and we note that as the value of $C_3$ increases from $C_{3}=0.99\,10^{-2}$GeV$^{3}$ , which corresponds to a $10\%$ contribution to the form factor at $Q_0$, to $2.98\,\ 10^{-2}$ GeV$^{3}$ , which corresponds to a $30\%$ contribution, again the value of the $\pi TFF$ decreases. Thus a small increase in $a$ and $C_3$ moves our result toward the Belle data.  Finally,  in Fig. \ref{Fig3}  we plot the better fit ($\chi^2/dof=1.21$) taking into account 
all the world data which corresponds to $a=0.032$ with the $C_3$ term at the $30\%$ value. We stress that there is no strong correlation between $a$ and $C_3$ as long as $a$ is kept within its experimental error bars. Thus the fit is quite stable with respect to the parameters of the low energy model.

The fit to the data is excellent  with a very small variation of the $1/Q^2$ contribution at $Q_0$ from previous fit, i.e. from $20\%$ to $30\%$. It must be said, before entering the discussion of this fit, that in our previous work \cite{Noguera:2010fe} we pointed out that the average value of the highest energy data points of BABAR were too large, a conclusion reached also by other analyses \cite{Mikhailov:2009sa,Dorokhov:2009zx}. In Fig. \ref{Fig3} we show not only the fit for $30\%$ contribution of $C_3/Q^2$ at $Q_0$, but its behavior for higher values of $Q^2$. As can be seen, also stressed in our previous work, this contribution is small in size. However, and this an important outcome of our analysis, it is instrumental in fixing the initial slope at the matching point, which  determines, after evolution, the high energy behavior of the form factor.  

In our opinion the Belle data confirm the BABAR result that the $\pi TFF$ crosses the asymptotic QCD limit. This limit is well founded under QCD assumptions, but nothing is known of how this limit is  reached, if from above or from below. BABAR and Belle data suggest that the limit is exceeded around  $10-15$ GeV. Our calculation is consistent with this result. The necessary growth of the $\pi TFF$ between $5-10$ GeV to achieve this crossing is in our case an indication of nonperturbative behavior and  $C_3/Q^2$ contribution at low virtuality. The determination of the crossing point is a challenge for any theoretical model and therefore, the precise experimental determination of it is of relevance. Many models fail to achieve this crossing because their pion DA is defined  close to its asymptotic form. 

The pion DA can be expressed as a series in the Gegenbauer polynomials,
\begin{equation}
\phi_{\pi}\left(  x,Q^{2}\right)  =6x\left(  1-x\right)  \left(
1+\sum_{n\left(  even\right)  =2}^{\infty}a_{n}\left(  Q^{2}\right)
\,C_{n}^{3/2}\left(  2x-1\right)  \right)
\end{equation}
We can compare different models by looking at the values of the coefficients 
of the expansion $a_{n}\left(Q^{2}\right)$. In our case, at $Q^{2}=1$ GeV$^{2}$ many 
$a_{n}$ coefficients are significant, but we focus our attention in a few
terms: $a_{2}=0.389,$ $a_{4}=0.244$ and $a_{6}=0.179$. At $Q^{2}
=4$ GeV$^{2}$ we obtain the values $a_{2}=0.307,$ $a_{4}=0.173$ and
$a_{6}=0.118$, which are close to those obtained by Polyakov \cite{Polyakov:2009je}.
Consistently, our result for the $\pi TFF$ is similar to that obtained in ref.
 \cite{Polyakov:2009je}. At $Q^{2}=5.76$ GeV$^{2}$ we obtain $a_{2}=0.292,$ $a_{4}=0.161$
and $a_{6}=0.108$, which are very different from those of ref.
\cite{Bakulev:2012nh}. These author use for their fit  BABAR data for the  $\eta TFF$ \cite{BABAR:2011ad}, together with the pion data. It  is therefore not a surprise
that these authors  come to a different conclusion, namely, that the Belle and the BABAR data cannot be reproduced to the same level of accuracy within the Light Cone Sum Rules approach \cite{Bakulev:2011rp}. 
However, in an extension of the ideas developed
in the present paper to the $\eta$ case  studied in ref. \cite{Noguera:2011fv}
looking at the state $\left\vert q\right\rangle =\frac{1}{2}\left(
\left\vert u\,\bar{u}\right\rangle +\left\vert d\,\bar{d}\right\rangle
\right)$  a  very different structure of the $a_{n}$ coefficients to that of the pion arises.
At $Q^{2}=1$ GeV$^{2},$ the values of the coefficients are $a_{2}=0.134$ and $a_{4}=0.352$  or,
equivalently, at $Q^{2} = 5.76$ GeV$^{2}$ we have$\ a_{2}=0.101$ and
$a_{4}=0.232.$ Therefore, that study does not supports the combined use of both
data sets

We have developed a formalism to describe the $\pi TFF$ on all experimentally accessible range, and hopefully beyond. The formalism is based on a two energy scale description. The formulation in the
low energy scale is nonperturbative, while that of the high energy scale is
based  on perturbative QCD. The two descriptions are matched at an energy scale $Q_{0}$  called hadronic scale \cite{Traini:1997jz,Noguera:2005cc}. We stress the crucial role played by the nonperturbative input at the level of the low energy description. It is an important outcome of 
this calculation the role played by the $1/Q^2$ power correction term in determining the slope of the data 
at high $Q^2$, despite the fact that they do almost not contribute to the value of the $\pi TFF$ .  

We have used a flat $\pi$ DA, i.e. a constant value for all $x$ \cite{Radyushkin:2009zg,Polyakov:2009je}, which with our normalization becomes $\phi(x)=1$. Our choice has been motivated by chiral symmetry \cite{Noguera:2010fe}.
Model calculations, Nambu-Jona-Lasinio (NJL) \cite{Anikin:1999cx,Praszalowicz:2001wy,RuizArriola:2002bp,Courtoy:2007vy} and the
"spectral" quark model \cite{RuizArriola:2003bs}, give a constant $\pi$ DA.  The $\pi TFF$ calculated in these models,
however, overshoots the data \cite{Broniowski:2009ft}, emphasizing the importance of QCD evolution.

The calculation shown proves that the BABAR and Belle results can be accommodated in our scheme, which only uses standard QCD
ingredients and low energy data. Moreover, at the light of our results, we confirm that  at $40$ GeV$^2$ we have not yet reached
the asymptotic regime which will happen at higher energies.

We would like to thank A. V. Pimikov and M. V. Polyakov for useful comments.This work has been partially funded by the Ministerio de Economía y Competitividad 
and  EU FEDER under contract FPA2010-21750-C02-01, by Consolider Ingenio 2010
CPAN (CSD2007-00042), by Generalitat Valenciana: Prometeo/2009/129, by the
European Integrated Infrastructure Initiative HadronPhysics3 (Grant number
283286).

\end{document}